\documentclass[12pt,dvips]{article}

\usepackage{epsfig}

\def\snu{{\langle \tilde{\nu} \rangle}}
\def\hu{{\langle H_u \rangle}}
\def\RMP#1#2#3{Rev. Mod. PHys.  {\bf #1}, #2 (19#3)}
\def\NPB#1#2#3{Nucl. Phys. {\bf B#1}, #2 (19#3)}
\def\PLB#1#2#3{Phys. Lett. {\bf B#1}, #2 (19#3)}

\def\PRD#1#2#3{Phys. Rev. {\bf D#1}, #2 (19#3)}

\def\PRL#1#2#3{Phys. Rev. Lett. {\bf #1}, #2 (19#3)}

\def\susy{{supersymmetry }}

\def\eV{{\,{\rm eV}}}
\def\GeV{{\,{\rm GeV}}}
\def\TeV{{\,{\rm TeV}}}

\def\snu{{\langle \tilde{\nu} \rangle}}
\def\hu{{\langle H_u \rangle}}
\def\hd{{\langle H_d \rangle}}
\def\nut{{\tilde{\nu}}}
\def\ep#1{\epsilon_{#1}}

\def\mt{{\tilde{m}^2}}
\def\huv{{\langle H^u \rangle}}

\def\centeron#1#2{{\setbox0=\hbox{#1}\setbox1=\hbox{#2}\ifdim
        \wd1>\wd0\kern.5\wd1\kern-.5\wd0\fi
        \copy0\kern-.5\wd0\kern-.5\wd1\copy1\ifdim\wd0>\wd1
        \kern.5\wd0\kern-.5\wd1\fi}}
\def\ltap{\;\centeron{\raise.35ex\hbox{$<$}}{\lower.65ex\hbox{$\sim$}}\;}
\def\gtap{\;\centeron{\raise.35ex\hbox{$>$}}{\lower.65ex\hbox{$\sim$}}\;}

\def\be{\begin{equation}}
\def\ee{\end{equation}}

\def\CO{{\cal O}}

\begin{document}

\begin{titlepage}
\begin{flushright}
UW/PT 99-02\\
%SCIPP?????\\
\end{flushright}

\vskip 1.2cm

\begin{center}
{\Large\bf Solar and Atmospheric  Neutrino Oscillations\\
 from Bilinear R-Parity Violation }

\vskip 1.4cm

{\large D. Elazzar Kaplan and Ann E. Nelson}
\vskip 0.4cm
{\it Department of Physics 1560, University of Washington,
                Seattle, WA 98195-1560\\}
\vskip 4pt

\vskip 1.5cm

\begin{abstract}
We discuss general predictions  for
neutrino masses and mixing angles from R parity violation in the 
Minimal Supersymmetric Standard Model. 
If the soft supersymmetry breaking terms are flavor blind at short distance, 
then the leptonic analogue of the CKM matrix depends on only two real 
parameters, which are completely determined by fits to solar and 
atmospheric neutrino oscillations.  Either the small angle MSW, large 
angle MSW, or ``just-so'' solutions to the solar neutrino problem are 
allowed, although the large angle MSW solution requires substantial 
fine-tuning.  The latter two cases require significant 
$\nu_\mu\rightarrow\nu_e$ oscillations of atmospheric neutrinos. 
We  present a model which could explain bilinear R parity violation 
as a consequence of spontaneous symmetry violation by a dynamical
supersymmetry breaking sector.  The decay length and branching ratios 
of the LSP are estimated.

\end{abstract}

\end{center}

\vskip 1.0cm

\end{titlepage}

\section{Evidence for neutrino oscillations and neutrino mass}

Recently Super-Kamiokande has reported  an energy dependent and zenith angle
dependent deficit of  atmospheric muon neutrino events compared with
theoretical expectations, which is strong evidence for 
oscillations of muon neutrinos \cite{superK}. The hypothesis of
neutrino oscillations  will soon be further tested through long baseline 
neutrino oscillation searches \cite{long}.  In addition, the solar 
electron neutrino flux observed by 5 different experiments  is much less 
than expected, suggesting oscillations of electron neutrinos as well
\cite{solarmod}. The simplest explanation for either anomaly is to modify
the Standard Model  to provide neutrinos with mass. 
The atmospheric neutrino data can be explained well by 
a large mixing angle  for the muon and tau neutrinos  and a neutrino 
mass squared difference between $\sim 5\times 10^{-4}$ and $\sim 6\times 
10^{-3}$~eV$^2$. The solar neutrino flux deficits can be explained by the 
MSW \cite{msw} effect for either a large or a small electron neutrino 
mixing angle and a mass squared difference of order 
$10^{-5}$~eV$^2$ \cite{mswsol}, or by ``just-so'' \cite{justso} vacuum 
oscillations with large electron neutrino mixing and a mass squared 
difference of $\sim10^{-10}$~eV$^2$.  If one relaxes the constraints 
coming from the standard solar model or from one or more of the solar 
neutrino experiments then other mass squared differences  and mixing 
angles are possible \cite{barbhall}.   

Oscillations amongst all three neutrinos can explain all solar and 
atmospheric data, provided one mass squared difference  is in the 
atmospheric range and another in  either the MSW or ``just-so'' 
solar ranges.  Some  models invoke oscillations to new neutrino 
states which are ``sterile'' under the weak interactions in order to 
also account for the $\nu_\mu\rightarrow\nu_e$ oscillation evidence 
reported by the LSND collaboration \cite{lsnd}.  Several future 
experiments on solar and terrestrially produced neutrinos are expected to 
further constrain the possibilities \cite{conrad}.

In this note we will assume that there are only three neutrino states
and will only explain the solar and atmospheric neutrino anomalies. We
neglect the results of LSND as these have not been confirmed by any
independent experiment\footnote{The KARMEN \cite{karmen} and
mini-BooNE \cite{miniboone} experiments have the capability to check
this result.}. We will show that a class of  renormalizable supersymmetric 
theories with lepton number violation can account for these anomalies 
and examine the consequent predictions.

\section{Bilinear R parity violation}

Unlike the minimal Standard Model, the Minimal Supersymmetric Standard Model 
(MSSM) allows renormalizable baryon and lepton number violation. If
neutrinos are massive Majorana particles, then lepton number is
violated by nature, and there is no compelling reason to assume that
lepton number violating terms are absent from the
superpotential\footnote{It is however trivial to find discrete
symmetries which forbid renormalizable lepton number violating terms 
but allow dimension five terms which give Majorana neutrino masses.}.

In the MSSM, the most general renormalizable superpotential  contains the terms
\be
\label{eq rps}
{1\over2}\lambda_{ijk}\ell_i\ell_j\bar{e}_k+\lambda'_{ijk}\ell_iq_j\bar{d}_k+
{1\over2}\lambda''_{ijk}\bar{u}_i\bar{d}_j\bar{d}_k \ .
\ee
Another possible term is
\be \label{eq bilinear}\epsilon_i H_u \ell_i\ee
but this is conventionally eliminated by performing a redefinition of the 
$H_d$ and lepton superfields
\be \label{eq trans}
\pmatrix{\ell_e\cr\ell_\mu\cr\ell_\tau\cr H_d}\rightarrow 
U\pmatrix{\ell'_e\cr\ell'_\mu\cr\ell'_\tau\cr H'_d}
\ee
where $U$ is a unitary  $4\times 4$ matrix.

To avoid proton decay, all such terms are usually forbidden by
imposing a discrete symmetry, known as R parity,  under which the 
quark and lepton  superfields change sign. However, avoiding 
proton decay only requires that either the baryon number violating terms 
or the lepton number violating terms vanish\footnote{This holds unless the 
model has a light non-leptonic fermion such as an axino or light gravitino.}. 
Supersymmetric models with either baryon or lepton number violation
are known as R parity violating models \cite{hallsuz,rpreview}.
A special case of R parity violation, known as  the bilinear R parity 
violating model (BRPVM),  only contains the R parity violating terms 
which can be written in the form (\ref{eq bilinear}). This restricted 
model is more predictive and theoretically more attractive than more 
general models.  The BRPVM arises naturally from a large class of 
theories in which R parity violation is spontaneous,   such as many  
grand unified models \cite{hallsuz}, and, unlike general R parity 
violating model, has few potential phenomenological difficulties
with flavor changing neutral currents and lepton flavor violation
\cite{nowakowski}.

\subsection{Superpotential and soft supersymmetry breaking terms}

Although the BRPVM is defined by assuming the superpotential can be written 
in the form (\ref{eq bilinear}), it is more convenient to use the 
transformation (\ref{eq trans}) to write the superpotential in the 
canonical form of (\ref{eq rps}). The lepton number  violating terms in 
the superpotential can then be written in terms of only 3 new parameters 
$\theta_i$, $i=1,2,3$ with
\be
s_i,c_i\equiv \sin\theta_i,\cos\theta_i\ee
and 
\be
s_1\equiv{\epsilon_e\over\sqrt{\epsilon_\mu^2+\epsilon_e^2}},\quad
s_2\equiv{\sqrt{\epsilon_\mu^2+\epsilon_e^2}\over\sqrt{\epsilon_\tau^2+
\epsilon_\mu^2+\epsilon_e^2}},\quad 
s_3\equiv {\sqrt{\epsilon_\tau^2+
\epsilon_\mu^2+\epsilon_e^2}\over\sqrt{\mu^2+\epsilon_\tau^2+
\epsilon_\mu^2+\epsilon_e^2}}\ ,
\ee where $\mu$ is the coefficient of the superpotential term $H_u H_d$.
Neglecting the Yukawa couplings of the first two generations,  the  
superpotential lepton number violating terms can be written
\be \label{eq trilinear}
\lambda_b{\bar b}q_3\{s_3[c_2\ell_\tau+s_2(c_1\ell_\mu+s_1\ell_e)]\}+
\lambda_\tau{\bar \tau}\ell_\tau\{s_3[s_2(c_1\ell_\mu+s_1\ell_e)]\}\ ,
\ee
where $q_3\equiv (t',b)_L$,  $t'$ is the electroweak partner of the $b$ 
quark, and $\lambda_{b,\tau}$ are respectively the bottom and tau Yukawa 
couplings. 

In general there will also be the soft supersymmetry breaking lepton number 
violating terms in the scalar potential
\be \label{eq nonuniversal}
B_{\epsilon_i}\epsilon_i H_u \tilde \ell_i + {\tilde m}^2_{H_d \ell_i}H_d^* 
\tilde \ell_i
+A^{\lambda}_{ijk} \lambda_{ijk} {\tilde{\bar d}}_i \tilde q_j \tilde\ell_k
+ A^{\lambda'}_{ijk} \lambda'_{ijk}{\tilde {\bar e}}_i\tilde 
\ell_j\tilde\ell_k
+h.c.
\ee
We  assume lepton universality for the  supersymmetry breaking terms, i.e.  
the messenger of supersymmetry breaking is blind to lepton flavor. 
This assumption allows us to simplify (\ref{eq nonuniversal}) to
\be
B_{\epsilon} \epsilon_i H_u \tilde \ell_i 
+ {\tilde m}^2_{H_d \ell}(\epsilon_i/\mu)H_d^* \tilde \ell_i
+A^{d}  \lambda^d_{i}(\epsilon_j/\mu){\tilde{\bar d}}_i \tilde q_i \tilde\ell_j
+ A^{e} (\lambda^e_{i}(\epsilon_j/\mu)-\lambda^e_{j}(\epsilon_i/\mu))\
{\tilde {\bar e}}_i\tilde \ell_i\tilde\ell_j
+h.c.
\ee
at the ``messenger scale'' $\Lambda$ at which supersymmetry breaking is 
communicated, although at another scale all the terms in 
(\ref{eq nonuniversal}) will be generated by renormalization.

Lepton number violation allows neutrinos to obtain  Majorana masses. 
One linear combination of neutrinos can gain a mass at tree level from the
diagram in Figure~\ref{heavy}a  and at one loop from Figure~\ref{heavy}b, 
while the loop diagrams of Figure~\ref{tauloopmass} can give the other 
neutrinos mass. The computations of sneutrino vevs contributing to these
masses may be found in the appendix.

Generally this model produces an enormous hierarchy in the neutrino mass 
spectrum, with the ratio of the heaviest and second heaviest neutrino 
masses at least $10^3$.  In order to  reduce this hierarchy, we can assume 
that the communication of supersymmetry breaking is also blind to the 
difference between a lepton and $H_d$, as would naturally occur with 
Gauge Mediated Supersymmetry Breaking \cite{gmsbrev} in which all soft 
supersymmetry breaking terms including those of (\ref{eq nonuniversal}) 
are generated by gauge interactions \cite{mmm}. In such models, one can 
choose a basis where the bilinear lepton number violating soft supersymmetry 
breaking terms  vanish at the messenger scale $\Lambda$, and loop effects 
are needed to generate a sneutrino vev.

\subsection{The lepton mixing matrix}

\begin{figure}
\begin{center}
        \epsfig{file = 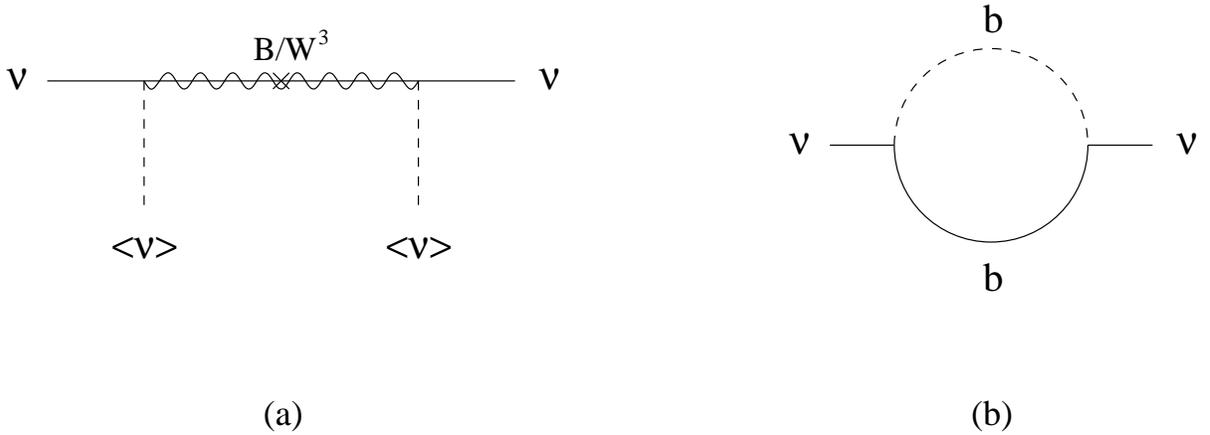}
        \caption{Contributions to the largest neutrino masses at 
		(a) tree level and (b) one loop.}
        \label{heavy}
\end{center}
\end{figure}

In the BRPVM, a very general argument shows that the lepton analogue of 
the CKM matrix can be predicted in terms of 2 mixing angles and is CP
conserving\footnote{Several recent papers \cite{rpneut} have analyzed 
neutrino masses in versions of the BRPVM, but have not noted this general 
prediction.}.  This is most easily seen in the basis where all R parity 
violation is bilinear. Note that in the excellent approximation of 
neglecting the electron and muon Yukawa couplings, there is a $U(3)$ flavor 
symmetry acting on the neutrinos which is broken by only two terms---the 
R parity violating term 
\be\label{rpterm}\sum_{i=e,\mu,\tau}\epsilon_iH_u\ell_i\ee
and the tau Yukawa coupling 
\be\label{tauyuk}\lambda_\tau \ell_\tau\bar \tau_3 H_d\ee 
(recall that we are assuming no flavor violation from  supersymmetry 
breaking). The linear combination \be(c_1\ell_e-s_1\ell_\mu)\ee
is invariant under a chiral $U(1)$ symmetry, which is broken only by tiny 
Yukawa couplings, and prevents this linear combination from gaining a mass.   
Thus one neutrino, which is purely a linear combination of $\nu_e$ and 
$\nu_\mu$, is always automatically very light compared with the other two. 
This argument is true for any of the possible mechanisms for generating the 
neutrino masses, provided only that the supersymmetry breaking terms respect
lepton universality. The heaviest neutrino mass has no suppression factor 
due to the tau Yukawa coupling. This neutrino mass  can result at 
tree level from sneutrino vevs,  as well as  from the one loop graph in 
Figure~\ref{heavy}.  Hence the heaviest neutrino, up to  corrections 
involving $\lambda_\tau$, is the linear combination 
\be\sum_{i=e,\mu,\tau}\epsilon_i\ell_i/\sqrt{\epsilon_\tau^2+\epsilon_\mu^2+\epsilon_e^2}=c_2\nu_\tau+s_2(c_1\nu_\mu+s_1\nu_e) 
\ .\ee 
The mass of the  second heaviest neutrino is proportional to both the 
R parity violating terms and the tau Yukawa coupling.

%Figure 1%

The preceding argument shows that in a basis where $j=1,2,3$ labels 
neutrino mass eigenstates in ascending order of mass,  the neutrino 
CKM matrix has the simple form
\be
V^\nu=\pmatrix{
V_{e1}&V_{e2}&V_{e3}\cr
V_{\mu1}&V_{\mu2}&V_{\mu3}\cr
V_{\tau 1}&V_{\tau2}&V_{\tau3}
}=\pmatrix{
c_1&s_1c_2&s_1s_2\cr
-s_1&c_1c_2&c_1s_2\cr
0&-s_2&c_2}\ .\ee
This is one of the main results of this paper.

For vacuum neutrino oscillations we predict
\begin{eqnarray}
P_{e\leftrightarrow \mu}&=& \sin^2 2\theta_1\left[
\left(c_2^2\sin^2{\Delta m_{12}^2 L\over 4 E}+
s_2^2\sin^2{\Delta m_{13}^2 L\over 4 E}\right)-{1\over4}
\sin^2 2\theta_2 \sin^2{\Delta m_{23}^2 L\over 4 E}\right]\cr
P_{e\leftrightarrow \tau}&=&s_1^2\sin^2 2\theta_2\sin^2{\Delta m_{23}^2
  L\over 4 E}\cr
P_{\mu\leftrightarrow \tau}
	& = 
	& c_1^2\sin^2 2\theta_2 
		\sin^2{\Delta m_{23}^2 L\over 4 E}.
\end{eqnarray}
Note in the limit where $\Delta m_{12}^2$ can be neglected
(relevant for atmospheric and terrestrial neutrino oscillations) that
$P_{e\leftrightarrow \mu}$    simplifies to
\be
P_{e\leftrightarrow \mu}\approx
s_2^4\sin^2 2\theta_1 \sin^2{\Delta m_{23}^2
  L\over 4 E}\ ,\ee 
while in the limit where $\Delta m_{23}^2$ is large
enough so that many oscillation lengths are averaged over (relevant
for ``just-so'' solar neutrino oscillations) 
\be
P_{e\not\rightarrow e}\approx
\sin^2 2\theta_1 c_2^2\sin^2{\Delta m_{12}^2
  L\over 4 E} +{2}s_1^2s_2^2(c_2^2+c_1^2s_2^2)\ .
\ee

In general, matter effects significantly change these results for both
solar and atmospheric neutrinos \cite{msw}. Several groups 
\cite{foglisol,fogliat,otherthree} have derived constraints on neutrino 
oscillation parameters, under the assumption that all 3 neutrino flavors 
mix. Using the results of Fogli et al. on atmospheric \cite{fogliat}  and 
MSW solar \cite{foglisol} neutrino oscillations and the results of ref.
\cite{threeso} on ``just-so'' solar neutrino oscillations, 
we find three regions of parameter space which can give a good fit to the 
SuperKamiokande data and all the solar neutrino experiments:
\begin{enumerate}
\item{} The angle $\theta_2$ is large ($\sin^2(2\theta_2)>0.8$) and 
$\theta_1$ is small ($s_1\sim 0.07$).  In this case the atmospheric 
neutrino oscillations are almost purely $\nu_\mu\leftrightarrow\nu_\tau$ 
and $m_3$ lies between $\sim3\times 10^{-2}$ and 
$ \sim10^{-1}$~eV. The solar neutrino problem is solved by the small angle 
MSW effect and $m_2\sim 3\times 10^{-3}$~eV. 

\item{} Both $\theta_2$ and $\theta_1$ are large, with $s_1\sim 0.7$ and 
$s_2\sim 0.8$, and the solar neutrino problem is solved by the large angle 
MSW effect. In this case $m_3\sim 3\times 10^{-2}$~eV  and 
$m_2\sim10^{-2}$~eV. Since the ratio $m_2/m_3$ is always rather small 
unless there are finely tuned accidental cancellations, this solution 
seems rather unlikely.

\item{} A small region of parameter space is within the 90\% confidence 
limits for both atmospheric neutrinos and ``just-so'' solar neutrinos with
$s_1\sim 0.5$,  $s_2\sim 0.8$, $m_2\sim 10^{-5}$~eV and  $m_3\sim 3\times 
10^{-2}$~eV.  This large ratio $m_3/m_2$ is easily produced in a generic
model of supersymmerty breaking.
\end{enumerate}

The long baseline reactor search for electron neutrino disapearance,  KamLAND
\cite{kamland}, will definitively distinguish these three possibilities, 
being sensitive to $\Delta m^2$ larger than $\sim~10^{-5}$~eV$^2$. 
In the small angle MSW  case there is a negligible possibility of $\nu_e$ 
disapearance,  the large angle case will give (averaging over both 
oscillation lengths) $P_{e\not\rightarrow e}\sim 0.6$ (note that an 
average disappearance probability greater than 1/2 is possible when the 
electron neutrino is a linear combination of more than two mass eigenstates) 
and in the just-so case, where only one $\Delta m^2$ is large enough to be
observable in a terrestrial experiment, $P_{e\not\rightarrow e}\sim 0.3$.

Note that in all cases, the ratio $m_2/m_3$ must lie between $\sim .3$
and $\sim 3\times 10^{-4}$. As we will discuss in the next section, $m_2$ 
is always proportional to a loop factor suppressed by powers of 
$\lambda_\tau$, and so models which can give a large enough value for
this ratio are quite constrained.

\subsection{Predictions for neutrino mass}

\begin{figure}
\begin{center}
        \epsfig{file = 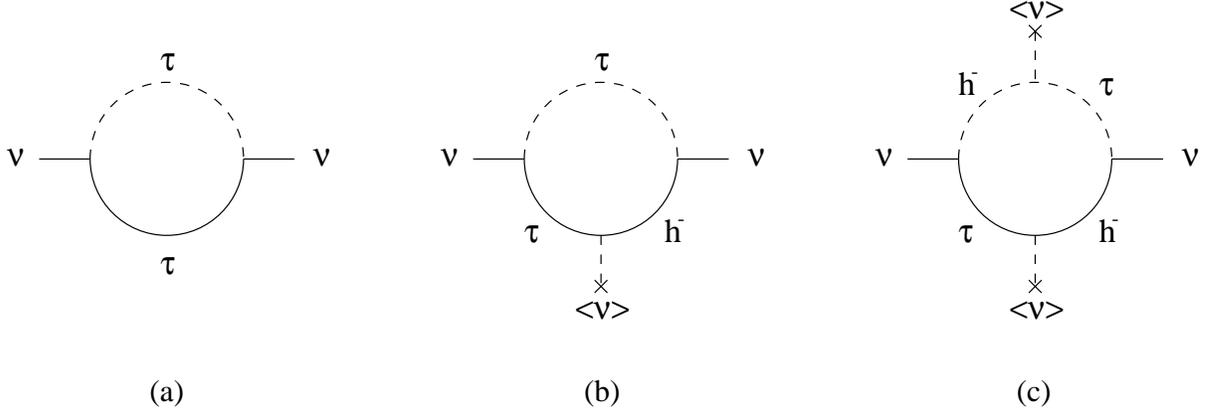}
        \caption{One loop contributions to the second neutrino mass.}
        \label{tauloopmass}
\end{center}
\end{figure}

The possible dominant contributions to the heaviest neutrino mass $m_3$ 
are from the tree diagram, Figure~\ref{heavy}a, and  the one loop diagram, 
Figure~\ref{heavy}b, while $m_2$ can only arise from the one loop 
diagrams\footnote{The diagrams in Figure~\ref{tauloopmass} have been effectively
taken into account in \cite{banks}, but in the basis where the neutrino vev
vanishes.  See also \cite{kiwoon}.}
of Figure~\ref{tauloopmass}, and is always suppressed by $\lambda_\tau$.   
Typically the tree diagram is much larger than any loop diagrams, and gives 
a contribution to $m_3$ of
\be\label{eq tree}
m_{3} = (\frac{g^2}{4 M_2} + \frac{g'^2}{4 M_1}) \snu^2,
\ee
where $g,g'$ are Standard Model gauge couplings, $M_{1,2}$ are Majorana 
gaugino masses and we have neglected mixing in the neutralino mass matrix 
(reasonable for $m_z\ll M_{1,2}$ or $m_z \ll \mu$).  However, the sneutrino 
vevs are naturally suppressed relative the the parameters $\ep{i}$, when 
universal soft supersymmetry breaking terms are generated at a low scale.
Note that small sneutrino vevs are necessary if $m_2/m_3$ is large enough
to allow the MSW solution to the solar neutrino problem and are most 
natural when $\hu/\hd\equiv\tan\beta$ is large, the soft \susy breaking
terms are completely universal, and the messenger scale is low, as in a
gauge mediated scenario.  In the appendix we compute the sneutrino vevs and 
find that a large value for $\tan\beta$, a messenger scale of $50\TeV$, 
and an accidental cancellation of order 10\% between two independent 
contributions to the sneutrino vev 
will allow the loop contributions to be within a factor of five of the 
tree contribution to $m_3$.  Thus we expect the tree contribution to 
dominate.  The contribution of the one loop diagram to $m_3$ is
\be\label{eq bloop}
m_{3}=\frac{\lambda_b^4 s_3^2}{16 \pi^2 \tan\beta} 
\frac{\mu \hu^2 \log(\frac{\tilde{m}_{b_1}^2}{\tilde{m}_{b_2}^2})}
{\tilde{m}_{b_1}^2 - \tilde{m}_{b_2}^2}\ ,
\ee
where 
\be
\tilde{m}_{b_{1,2}}^2 = {1\over 2}\left[ 
\left(\tilde{m}_{b_L}^2 + \tilde{m}_{b_R}^2 \right) \pm 
\sqrt{\left(\tilde{m}_{b_L}^2 - \tilde{m}_{b_R}^2 \right)^2 + 
4 \lambda_b^2 \mu^2 \hu^2} \right]\ .
\ee

%Figure 2%

There are several possible contributions\footnote{These contributions
are not strictly to $m_2$ but to a linear combination of $m_2$ and $m_3$.
For $s_2^2\sim {1\over2}$ (approximately true in all cases), there is an 
additional contribution to the second heaviest neutrino's mass of order
$m_2^2/m_3$, and to the mixing of order $m_2/m_3$.  These contributions
are only non-negligiable in the MSW scenario where $m_2/m_3\sim 
{1\over 10} - {1\over 30}$.}
to $m_2$---which one is
largest depends on the size of $s_3$.  For large $s_3$,
the diagram of Figure~\ref{tauloopmass}a dominates, giving
\begin{equation}\label{eq tauloop}
m_{2}=\frac{\lambda_{\tau}^4 s_3^2 s_2^2 c_2^2}{16 \pi^2 \tan\beta} 
\frac{\mu \hu^2 \log(\frac{\tilde{m}_{\tau_1}^2}{\tilde{m}_{\tau_2}^2})}
{\tilde{m}_{\tau_1}^2 - \tilde{m}_{\tau_2}^2}\ ,
\end{equation}
where  $\tilde{m}_{\tau_{1,2}}$ are defined analogously to
$\tilde{m}_{b_{1,2}}$. 

For moderate $s_3$ and large $\tan\beta$, the dominant contribution to 
$m_2$ would come from the diagram of Figure~\ref{tauloopmass}b, giving
\be\label{eq onevevtau}
m_{2}= \frac{\lambda_{\tau}^4 s_3 s_2^2 c_2^2}{16 \pi^2} \frac{\snu \hu}{\mu}
f(\frac{\tilde{m}_{\tau_1}^2}{\mu^2},\frac{\tilde{m}_{\tau_2}^2}{\mu^2})\ ,
\ee
where
\be
f(x,y)=\frac{x \log(x) - y \log(y) - x y \log({x\over y})}{(1-x)(1-y)(y-x)}\ .
\ee
When this diagram dominates the ratio $m_2/m_3$ is always too 
small for an MSW solution to the solar neutrino problem, however a 
``just-so'' solution is possible for large $\tan\beta$.

Finally, for small $s_3$, the dominant contribution to $m_2$
comes from the diagram of Figure~\ref{tauloopmass}c, which gives
\begin{equation}
m_{2}= \frac{\lambda_{\tau}^4}{16 \pi^2} \frac{A_{\tau} \snu^2}{\mu^2} 
\left[f(\frac{m_{h^{-}}^2}{\mu^2},\frac{\tilde{m}_{\tau_1}^2}{\mu^2})
s_{\theta_\tau}^2 +
f(\frac{m_{h^{-}}^2}{\mu^2},\frac{\tilde{m}_{\tau_2}^2}{\mu^2})
c_{\theta_\tau}^2\right]\ ,
\end{equation}
where $\theta_\tau$ is the mixing angle between the left and right handed
tau sleptons.  Note that this contribution is very small unless 
$\lambda_{\tau}$ is large, which requires large $\tan\beta$. Large 
$\tan\beta$ is possible only if either $h_d$ is heavy or the soft 
supersymmetry breaking bilinear $h_uh_d$ term is small. Known mechanisms 
for suppressing such a bilinear term suppress the trilinear $A_{\tau}$ 
term as well. Thus for this case, for any known predictive  theory,
$m_2/m_3$ is  too small to allow a simultaneous fit to atmospheric and 
solar neutrino data.

Note that the ratio $m_2/m_3$ is proportional to $\lambda_{\tau}^4$ in all
cases.  Thus obtaining a large enough value of $m_2/m_3$ to account for both 
solar and atmospheric neutrino anomalies always requires large  
$(\sim  50)\  \tan\beta$.

One additional way in which in may be possible to produce a viable
hierarchy of neutrino masses is through gravitational effects.  In the 
case of a large vev compared 
to $\epsilon\equiv\sqrt{\ep{\tau}^2 + \ep{\mu}^2 +\ep{e}^2}$, $m_2$ is too 
small even for the ``just-so'' solution.  However, $m_2$ may receive an 
interesting contribution from a nonrenormalizable operator in the 
superpotential:
\be
W\supset \lambda^{\nu}_{ij} \frac{H_u \ell_i H_u \ell_j}{M_p}
\ee
where $M_p\simeq 2\times10^{18}$ is the reduced Planck mass.  With 
$\lambda^{\nu}$ of order one, this produces a neutrino mass 
$m_{\nu}\sim \huv^2/M_p \sim 10^{-5}\eV$, precisely what is needed for 
``just-so'' solar neutrino oscillations.  However, there is no definite
prediction for the leptonic CKM mixing matrix in this case.

\section{Models of bilinear R parity violation}

Supersymmetric models which generate only bilinear R parity violation
are quite natural.  One way to produce such terms is via a spontaneously
broken R symmetry.  For example, a U(1)$_R$ symmetry exists in the MSSM 
with the following charge assignments\footnote{A continuous U(1)$_R$
symmetry has SU(2)$\times$U(1) anomalies.  However, R parity is an anomaly 
free discrete subgroup.  The anomalies could always be cancelled by another
sector, such as a messenger sector.}:
\newline
\vskip .1in
\begin{center}
\begin{tabular}{|c|c|c|c|c|c|c|c|}\hline
Superfield& $q$ & $u^c$ & $d^c$ & $\ell$ & $e^c$ & $H_u$ & $H_d$\\
\hline
\raisebox{-2pt}{U(1)$_R$}& \raisebox{-2pt}{${1\over 2}+a$} 
& \raisebox{-2pt}{${1\over 2}-a$} & \raisebox{-2pt}{${1\over 2}-a$} 
& \raisebox{-2pt}{${1\over 2}+b$} & \raisebox{-2pt}{${1\over 2}-b$} 
& \raisebox{-2pt}{$+1$} & \raisebox{-2pt}{$+1$} \\[4pt]
\hline
\end{tabular}
\end{center}
\vskip .15in
\noindent For $a=-{b\over 3}$, all renormalizable R parity violating terms 
have R charge (${3\over 2}+b$).

Now, at a scale $M$, we introduce a new sector of particles which 
does not couple to the observable sector via contact or gauge interactions,
and in which the U(1)$_R$ symmetry is broken spontaneously.  Thus the 
R symmetry breaking can only be communicated via gravitational effects.  
Assume there exists a gauge invariant operator $\CO$ of mass 
dimension $n$ and R charge $({1\over 2}-b)$.  If $\CO$ has an expectation 
value $\langle\CO\rangle=M^n$, then the following term in the superpotential
of the low energy effective theory\footnote{In general, there will be a
number of gauge invariant operators which fit this description.  Here we
consider the operator with the lowest mass dimension $n$.  Those operators 
with mass dimension greater than $n$ will have contributions suppressed by
powers of the Planck mass.},
\be
\frac{\CO}{M_p^{n-1}}\ell H_u,
\ee
produces a bilinear R parity violating term with coupling 
$\epsilon\sim M^n/M_p^{n-1}$.  Trilinear R parity violating terms would in 
general also be produced with couplings of order $M^n/M_p^n$, and therefore
would be greatly suppressed compared to the bilinear term.

An interesting possibility is to have the operator $\CO$ come from the
\susy breaking sector.  However, in order to produce a neutrino spectrum 
from the R parity violating term which is consistent with both solar and
atmospheric anomalies, we have seen that the parameter $\epsilon$ must 
not be much smaller than the sneutrino vev $\snu$.  This naturally occurs 
if all lepton violating soft breaking terms in (\ref{eq nonuniversal}) 
have coefficients no larger than $B \epsilon$, with $B$ of order the weak 
scale.  However, if $\CO$ is in the supersymmetry breaking sector, it 
would generally also have an F-term expectation value leading to a scalar 
bilinear of order
\be
\frac{F M^{n-1}}{M_p^{n-1}} {\tilde\ell} H_u.
\ee
Generically, the coefficient of this term is much too large.  For example, 
if $F/M \sim 10^5\GeV$ (as is typical in models of gauge mediated 
supersymmetry breaking) then $\snu/\epsilon \sim 10^2$.  Therefore, if 
R~symmetry breaking and supersymmetry breaking come from the same sector, 
then the heaviest neutrino's mass $m_3$ comes from R~parity violation, while 
$m_2$ comes from Planck slop, as described at the end of Section 2.3.

\section{Predictions for the decays of the (N)LSP}

In the BRPVM, with parameters chosen to fit the atmospheric and solar
neutrino data, most effects of the lepton number violating couplings are 
too small too be observed. The main evidence for R parity violation will 
only come when supersymmetry is discovered.  The Lightest Supersymmetric 
Particle (LSP) decays via the R parity violating couplings. In some
models, the LSP is the an ultralight gravitino, however the lifetime for 
the decay of the next lightest supersymmetric particle (NLSP) into the 
gravitino is very long, and so the NLSP will primarily decay via R parity 
violating couplings. 

How the (N)LSP decays depends on whether neutrinos oscillate according
to the MSW solution or the ``just-so'' solution.  The MSW case requires
a small sneutrino vev compared to the R parity violating parameters,
$\snu\ltap {\epsilon\over 100}$, in order for (\ref{eq tree}) to give a 
mass which satisfies atmospheric neutrino data.  Decays which depend on 
couplings proportional to $s_3 \simeq {\epsilon\over\mu}$ dominate.  The 
``just-so'' solution requires $\snu\gtap\epsilon$ to give a large mass
ratio $m_3 / m_2$.  In this case, the dominant decays of the (N)LSP are
due primarily to the mixing of leptons with neutralinos and 
charginos (except in decays to top quarks as mentioned below).
With the R parity violating couplings fixed to explain neutrino masses 
and mixing, we can make definite predictions for the dominant decay 
mode of the (N)LSP, which depends only on which superpartner it is and 
its mass.  In some cases, we can also predict the lifetime.

For instance, in the MSW case, if the (N)LSP is the lightest tau slepton 
$\tilde\tau_1$ (and ${\tilde m}_{\tau_1} < m_{\mbox{\tiny top}}$) it will 
decay into a charged lepton and neutrino with a lifetime between $10^{-15}$ 
and $10^{-16}$ seconds.  The $\tilde\tau_1$ could decay somewhat less often 
into $b\bar c$ quark jets.  The branching ratios of $\tau\nu$, $\mu\nu$, 
$e\nu$ and $b\bar c$ are proportional to $(\lambda_{\tau} s_2)^2$, 
$(\lambda_{\tau} s_2 c_1 s_{\theta_{\tau}})^2$, 
$(\lambda_{\tau} s_2 s_1 s_{\theta_{\tau}})^2$ and  
$3(\lambda_b V_{cb} c_2 c_{\theta_{\tau}})^2$ respectively.  In the case of 
the ``just-so'' solution, the $\tilde\tau_1$ decays nearly exclusively into 
$\ell\nu$ (mostly $\tau\nu$) with a lifetime of about $10^{-13}$ seconds.  
However, if ${\tilde m}_{\tau_1}\rightarrow b\bar t$ is kinematically allowed, 
it will be the dominant decay mode in the MSW scenario and may be comparable 
to the leading decay mode in the ``just-so'' case.

Another likely possibility for the (N)LSP is the lightest neutralino $N_1$, 
which has a small neutrino component. In the MSW scenario, 
the $N_1$ will decay predominantly into $b\bar b$ quark jets and a neutrino
(or $b\bar t\bar\ell$ if energetically allowed).  The lifetime in this case is
proportional to four powers of the mass $\tilde m_{\ell}$ of the exchanged
slepton when $\tilde m_{\ell}^2\gg M_{N_1}^2$.  In the case of the ``just-so''
solution, the dominant decay will be a charged lepton and a $W^\pm$,  with 
an interesting decay length of 0.1-10 mm~\cite{collider}.  The branching 
ratios for  $N_1$ decays into $e$, $\mu$ and $\tau$ are respectively 
proportional to $s_2^2s_1^2$, $s_2^2c_1^2$, and $c_2^2$. Also possible are 
decays into a neutrino and a $Z$, with a branching fraction approaching  
25\% when the $N_1$ is very heavy and mostly Bino \cite{collider}.  There
can also be a significant branching ratio for the mode $h^0 \nu$ if 
kinematically allowed.

\section{Summary}

The Bilinear R parity Violating Model is a well motivated and predictive 
theory  of neutrino masses.  The BRPVM can account for both the 
atmospheric and solar neutrino anomalies, and restricts the  form of the 
leptonic charged current mixing matrix, which can depend only on two 
independent angles. A large electron neutrino mixing angle solution to
the solar neutrino problem is possible if and only if the atmospheric
oscillations involve a significant $\nu_e$ component. The hierarchy
between neutrino masses is generically very large, and the MSW solution to 
the solar neutrino problem is reasonably natural only if $\tan\beta$ is 
large, soft supersymmetry breaking terms for the leptons and Higgses are
universal, and the scale of transmission of supersymmetry breaking is
low, suggesting gauge mediated generation of the soft supersymmetry
breaking terms.  With the typical neutrino mass ratio, vacuum oscillations
can solve the solar neutrino problem.  Definitive tests of this theory will
be provided by KamLAND which can distinguish the MSW from ``just-so''
solutions, and by observation of the pattern of (N)LSP decays.

\section*{Acknowledgements}

This work was supported in part by the Department of Energy Grant
No. DE-FG03-96ER40956.  DEK thanks J.~Gray and AN and DEK thank the UCSC 
particle theory group for their hospitality during the completion of 
this work.

\section*{Appendix: on sneutrino vev diagrams}

Some suppression of sneutrino vevs is necessary if the heaviest
neutrino is not to be too much heavier than the next heaviest. Here we
examine the sneutrino vevs assuming that at some scale $\Lambda$ there 
is no supersymmetry breaking mixing term between Higgs scalars and 
sneutrinos. Motivated by gauge mediated models,  we also assume the $A$ terms
vanish at $\Lambda$. We assume that $\Lambda$ is small enough so that we do 
not need to use the renormalization group, since when $\Lambda$ is large 
so will be the sneutrino vev. 

At one loop, there are contributions to $\nut$ - $H_d$ mixing:
\begin{equation}
{\tilde m}_{H_d \ell_i}^2 
	= {\mt}_{H_d \ell} \frac{\epsilon_i}{\mu} 
	= \frac{N_c \lambda_b^2 \epsilon_i}{16 \pi^2 \mu} 
        \left({\mt}_{b_1} \log{\frac{\Lambda^2}{{\mt}_{b_1}}} + 
	{\mt}_{b_2} \log{\frac{\Lambda^2}{{\mt}_{b_2}}}
	\right).
\end{equation}
There is also a  one loop  contribution to the $A$ terms  at scale $k$:  
 
\begin{equation}
A_b (k^2) = \left(4\over 3 \right) \frac{\alpha_s}{2 \pi} \lambda_b s_3 M_3 
        \int_0^1 \log{ \frac{\Lambda^2}{x\left( (1-x) k^2 + M_3^2 \right)}}dx
\end{equation}
where $M_3$ is the gluino mass.  The graphs with winos/binos are somewhat 
smaller, for example the wino contribution will be suppressed compared to 
the above by only $\sim~\left({3\over 4} \alpha_2 M_2 \right)/ 
        \left( {4\over 3} \alpha_3 M_3 \right) \sim {1\over 10}$.
The one loop contribution to $\nut$ - $H_u$ mixing is
\begin{equation}
B_{\epsilon_i} \epsilon_i = B_{\epsilon} \epsilon_i 
	= \left(4\over 3 \right) \frac{\alpha_s}{2 \pi} 
        \frac{\lambda_b^2}{16 \pi^2} \frac{\epsilon_i}{\mu} N_c \mu M_3
        \left[ 
        - {1\over 2} \left(\log{\frac{\Lambda^2}{\mt_b}}\right)^2
        - \log{\frac{\Lambda^2}{\mt_b}} 
        \right],
\end{equation}
where we've assumed small chargino mixing.  For simplicity, the result
shown is to zeroth order in $(\mt_{b_2} - \mt_{b_1})$.

The higgs vevs give us  terms  in the potential which are linear in 
the sneutrinos with a coefficient
\begin{equation}
C_{\nut} = \left( B_{\epsilon} \mu +
        \frac{{\tilde m}_{H_d \ell}^2}{\tan{\beta}}\right)
	s_3 \langle H_u \rangle,
\end{equation}
and thus give non-zero sneutrino vevs.  For degenerate sneutrino masses, 
the vev is $\langle \nut_3 \rangle = \frac{C_{\nut}}{\mt_{\nu}}$,
where $\nut_3$ is the partner of $\nu_3$, the heaviest neutrino in our basis.  
The sneutrinos are not exactly degenerate since $\mt_{\nu_{\tau}}$ will get 
an additional contribution from one-loop diagrams proportional to 
$\lambda_{\tau}^2$, but  this effect is small.

\end{document}